\documentclass[11 pt]{article}
\usepackage[dvips]{graphicx}

\input{tcilatex}
\begin{document}

\title{The wetting problem of fluids on solid surfaces. Part 2: the contact angle
hysteresis}
\author{Henri GOUIN\thanks{%
E-mail:henri.gouin@univ.u-3mrs.fr, telephone: +33 491288407, fax:
+33 491288776}}
\date{Laboratoire de Mod{\'e}lisation en M{\'e}canique et
Thermodynamique, EA2596, Universit{\'e} d'Aix-Marseille, 13397
Marseille Cedex 20, France} \maketitle \centerline {{\it
Communicated by\ } Kolumban Hutter, {\it Darmstadt}}

\begin{abstract}
In part 1  (Gouin, \cite{gouin}), we proposed a model of dynamics of
wetting for slow movements near a contact line formed at the
interface of two immiscible fluids and a solid when viscous
dissipation remains bounded. The contact line is not a material line
and a Young-Dupr\'e equation for the apparent dynamic contact angle
taking into account the line celerity was proposed.
\newline In this paper we consider a form of the interfacial
energy of a solid surface in which many small oscillations are
superposed on a slowly varying function. For a capillary tube, a
scaling analysis of the microscopic law associated with the
Young-Dupr\'e dynamic equation yields a macroscopic equation for the
motion of the contact line. The value of the deduced apparent
dynamic contact angle yields for the average response of the line
motion a phenomenon akin to the  stick-slip motion of the contact
line on the solid wall. The contact angle hysteresis phenomenon and
the modelling of experimentally well-known results expressing the
dependence of the apparent dynamic contact angle on the celerity of
the line are obtained. Furthermore,  a qualitative explanation of
the maximum speed of wetting (and dewetting) can be given.
\end{abstract}


\section{Introduction}

In \cite{gouin}, henceforth referred to as  part 1, we proposed a
model of non-Newtonian fluids for slow movements in the immediate
vicinity of the contact line that is formed at the interface of two
immiscible fluids and a solid. Fluid interfaces are modelled by
differentiable manifolds endowed with constant capillary energy.
Solid surfaces are also regarded as differentiable manifolds endowed
with a position dependant surface energy (we only consider the case
of surfaces without surfactant). The kinematics of slow isothermal
movements close to the contact line, revisited in the framework of
continuum mechanics, required the adherence condition to be relaxed
at the contact line. The velocity field is discontinuous at the
contact line and generates a concept of line friction but viscous
dissipation remains bounded.

\noindent Simple observations associated with the motion of two
fluids in contact with a solid wall reveal the following behaviour:
depending
 upon whether the fluid on the wall advances or retreats, a
variable contact angle is observed. The value of this apparent
dynamic contact angle, also called simply{\it Young's angle},
depends upon the contact line celerity. These observations resolve
distances which are not shorter than a few microns.
\medskip

The most notable unanswered problem is connected with the equation
controlling the macroscopic motion of the contact line and the
justification of contact-angle hysteresis  when the fluid is
advancing or receding on a solid wall.\newline To verify the
accuracy of the model proposed in part 1 and, particularly to
justify the Young-Dupr\'{e} equation for the apparent dynamic
contact angle associated with the notion of line friction, a simple
academic example is considered for which the equations of motion
together with the boundary conditions can be integrated in a
suitable approximation. This is the case of a thin cylinder
containing an incompressible fluid separated from air by a meniscus.
The apparatus is a capillary test-tube with rotation of symmetry.
In our example, we assume that
 the solid wall of the tube is endowed with a fluid-solid surface
energy which oscillates periodically with small variations with very
short wave length relative to the length of the
tube\footnote{Surface energies of solid walls with many small
wiggles, arising from small-scale microstructural changes, appear
often in scientific problems; for example, phase transformations,
protein folding and friction problems (Abeyaratne, Chu and James,
\cite{abeyaratne2}). The scale of microstructural changes is a
microscopic one and consequently is of an order smaller than the
length of the tube.}. The fact that we have chosen oscillations
superposed on a slowly varying function may be justified by the
possibility to solve the equations of motion analytically. More
general forms could also be considered.

\noindent For suitable dimensionless numbers corresponding to
\textit{slow} movements, the equations of motion of the liquid and
conditions on interfaces take simplified forms. The two-fluid
interface can then be modelled as a  spherical cap. It turns out
that \textit{the microscopic motion} of the contact system is
governed by a simple differential equation. This equation is
analytically decoupled from those of the liquid motion. When the
irregularities of the liquid-wall-surface energy vary over a length
that is vanishingly small relative to the size of the capillary
tube, the solution of the microscopic motion tends to a limit which
is the solution of a new differential equation. This differential
equation of \textit{the macroscopic or homogenized} motion is not
the limit of the microscopic equation.

\noindent The liquid in the capillary tube is controlled by a
piston. We deduce the motion of the two-fluid interface and we study
the behaviour of the apparent dynamic contact angle for the advance
and retreat of the contact line. In so doing a hysteresis phenomenon
appears. The deduced results are compared with those obtained in the
literature  with methods of statistical physics and experimental
measurements.

\section{The capillary tube apparatus}

The apparatus is a cylindrical tube of radius $\,a\,$ with a
vertical symmetry axis $0\,\mathbf{k}$. A liquid of volume $V_{0}$
fills the cylinder above a position determined by a piston. In
accordance with the hypotheses and notations of part 1, the liquid
is in contact with air through an interface $ \Sigma _{2t}$; the
constant air-liquid-surface energy is $\sigma _{2}\equiv \sigma
_{AB}$. In the motion, air is considered as incompressible. The wall
of the cylinder (piston included) is denoted by $\Sigma _{1t}$. The
wall is inhomogeneous and the surface energy $\sigma _{1}\equiv
\sigma _{AS}-\sigma _{BS}$ (difference between the superficial
energies of solid-liquid and solid-air) depends on the geometrical
position on $\Sigma _{1t}$. The value $\sigma _{1}$ is assumed to be
rotationally invariant. The
contact line is the curve $\Gamma _{t}$ connecting the two interfaces $%
\Sigma _{1t}$ and $\Sigma _{2t}$. Due to the axi-symmetric geometry,
its representation in fig. 1 is a point $P$ of which the position is
given by the abscissa $z$; the point $J$ of the piston,   the
position of which is given by the abscissa  $L$, is a function of
time commanded by an operator.

\begin{figure}[h]
\begin{center}
\includegraphics[height= 7.5 cm, width= 5 cm]{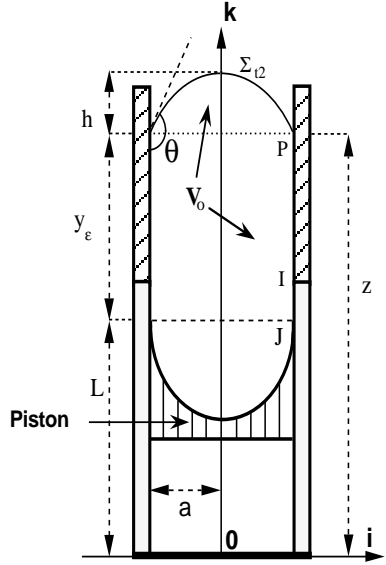}
\end{center}
\caption{Circular cylinder with vertical axis filled with a liquid
of volume $V_0$ held in position by a piston having cap form. The
part of the cylinder wall with the constant surface energy
$\sigma_{10}$ is colored in grey; the part of the cylinder wall with
the non-uniform surface energy $\sigma_{1}$ is hachured (see main
text). Referred to the orthonormal axis
$0\,\mathbf{i}\,\mathbf{k}\,$, $z$ denotes the position of the
contact line, $L=L(t)$ determines the position of the piston
$(z=y_{\protect\varepsilon }+L)$. The meniscus at the upper free
surface is rotational symmetric; the apparent dynamic contact angle
between the meniscus and the wall is denoted by $\protect\theta $;
$h$ is the  height of the meniscus. } \label{fig1}
\end{figure}
\noindent On the interval $JI$ of the wall, the surface energy
$\sigma_1$ is assumed to be constant with the value $\sigma_{10}$.
We consider the   case when the surface energy $\sigma_1$ on the
interval $IP$ of the wall is such that
$$
\sigma_1 = \sigma_{10}\, (\, 1 - k\, \mathrm{sin}\, {\frac{z}{\varepsilon\, a%
}}\, ), \eqno (1)
$$
where $0 < \varepsilon \ll 1$   is a small dimensionless parameter,
$\mid k \mid < 1 $  and $k\, \sigma_{10}$ is positive.
\newline Let us note that for sufficiently small  $\varepsilon$, the
average value of $
\sigma_1$ on any unspecified macroscopic set of the wall tends to $%
\sigma_{10}$ when $\varepsilon$ tends to zero.   In relation (1) the
interfacial energy $\sigma_1$ is built by two terms. The second is $
f_\varepsilon = - k \ \sigma_{10} \ \mathrm{sin}\, \Big ( z /(
\varepsilon\, a)\Big ) \, $. The distribution $f_\varepsilon $
converges to zero when $\varepsilon$ tends to zero and its virtual
work on $\Sigma_{t1}$ is null\footnote{ Let us calculate the value
of the distribution $ g_\varepsilon = \mathrm{sin}
\,(z/\varepsilon)$ when $\varepsilon \rightarrow 0, \, (\varepsilon
> 0)$. For any
function $\phi \in C^\infty [R,R]$ with compact support,\\
$ <g_\varepsilon , \phi > = \int_{-\infty}^{+\infty} \mathrm{sin} \,
(x/\varepsilon)\, \phi (x)\, dx = \varepsilon \,
\int_{-\infty}^{+\infty} \mathrm{sin} \, v\, \phi (\varepsilon v)\,
dv =
 -\varepsilon   \, [\,
\mathrm{cos}\, v \, \phi (\varepsilon v) \,]_{-\infty}^{+\infty} +
\varepsilon^2 \, \int_{-\infty}^{+\infty} \mathrm{cos} \, v\,
\phi^{\prime}(\varepsilon v)\, dv $.
\par
\noindent The first term $[\mathrm{cos}\, v \, \phi (\varepsilon v)
]_{-\infty}^{+\infty} = 0$ and $\mid <g_\varepsilon , \phi > \mid \,
\leq \, \varepsilon^2 \int_{-\infty}^{+\infty} \mid
\phi^{\prime}(\varepsilon v) \mid dv = \varepsilon
\int_{-\infty}^{+\infty} \mid \phi^{\prime}(x) \mid dx$ . From
$\int_{-\infty}^{+\infty} \mid \phi^{\prime}(x) \mid dx < \infty$,
we obtain $\displaystyle
\lim_{\varepsilon\rightarrow 0^+}\, <g_\varepsilon , \phi >\, = 0$ and $%
\displaystyle \lim_{\varepsilon\rightarrow 0^+} g_\varepsilon = 0$
in a weak sense.}. We understand that the associated force has an
action only on the contact line $\Gamma_t$.

\noindent It is proposed to study the motion of the contact line,
i.e. to determine the value of the position $z(t)$ as a function of
time. To this end, it is necessary to study the liquid motion.

\section{Equations of motion and boundary conditions}

In part 1, we proposed the equations of motion and boundary
conditions of a capillary motion of two fluids in contact with a
solid wall.
 The equations of motion are
\[
\rho a_{i}+p_{,i}=\phi _{i}+Q_{i,j}^{j}\, ,
\]
where ${\mbox{{\boldmath $\phi$}}}$ denotes the volumetric force,
$\rho $ the density, $\mathbf{a}$ the acceleration vector,
$\mathbf{Q}$ the (non-Newtonian) viscous stress tensor and $p$ the
pressure. At the solid wall, adherence conditions are required
except at the contact line.  On the meniscus $\Sigma _{t2}$, the
boundary condition is
$$
{\frac{{2\sigma _{AB}}}{R_{m}}}\ n_{2i}=(Q_{Ai}^{j}-Q_{Bi}^{j})\
n_{2j}+(p_{B}-p_{A})\ n_{2i}\, ,\eqno(2)
$$
where the indices $A$ and $B$ refer to the fluids $L_A$ and $L_B$,
$\mathbf{n}_{2}$ is the unit vector  of $\Sigma _{t2}$ external to
$L_B$,  and $R_{m}^{-1}$ is the mean curvature of $\Sigma
_{t2}$.\newline The boundary conditions on $\Sigma _{t1}$ introduce
an additional unknown scalar expressing the action of the solid wall
on the fluids (see Eq. (7), in part 1).\newline The only pertinent
parameter in the vicinity of the contact line is the {\it apparent
dynamic contact angle}, formulated as an implicit function of the
contact line celerity (see Section 5, in part 1); the {\it intrinsic
contact angle} (see Eq. (10), in part 1) does no longer appear  in
our  continuum mechanics point of view. Consequently, we denote
simply by $\theta $   the apparent dynamic contact angle and, in the
following, we  refer to it as the \textit{Young angle}. The dynamic
Young-Dupr\'{e} equation on the contact line is
$$
\sigma _{2}\ \mathrm{cos}\ \theta +\ \sigma _{1}+\nu \ u=0\eqno(3)
$$
where $\nu $ is the {\it line friction} and $u$ the contact line
celerity. The functional representation of $\nu $ and its value were
studied for a plane two-dimensional motion in part 1 and can be
extended to an axi-symmetric motion. The line friction depends on
the apparent dynamic contact angle (see Eq. (27) in part 1), and is
positive.
\newline
It is the problem of the meniscus that determines the moving
boundary including the contact line. To this end, a complete
picturing of the fluid-fluid interface is necessary. In fact, the
geometry of the meniscus is a consequence of the flow of the two
fluids in the vicinity of the meniscus and this flow is a functional
of pertinent dimensionless parameters. Without wanting to redo what
many authors have previously done, it is useful to recall the main
results known in the literature.

\noindent Many papers are concerned with the shape of the meniscus
near the contact line in capillary tubes. West, \cite{west}, was
among the first to study  this problem. Concus, \cite{concus},
presented an analysis of the static meniscus in a right cylinder.
Analysis using the Navier-Stokes equations and comparing results
with experiments of advancing interfaces in cylinders were
undertaken by Hoffman, \cite{hoffman}, Legait and Sourieau,
\cite{legait}, Finlow, Kota and Bose, \cite{finlow}, Ram\'e and
Garoff, \cite{rame}. Zhou and Sheng, \cite{Zhou}, analyzed the link
between macroscopic behaviour of the displacement of immiscible
fluids in a capillary tube and the microscopic parameters governing
the dynamics of the moving contact line; Thompson and Robbins,
\cite{thompson}, performed molecular dynamics simulations in the
hydrodynamics of the contact line; Dussan, Ram\'e and Garoff,
\cite{dussan1}, Decker {\it et al}, \cite{decker}, gave
considerations on the contact angle measurements; Voinov,
\cite{voinov}, proposed a thermodynamics approach to the motion of
the contact line. In all these papers, the following dimensionless
numbers are of significance,\newline

 Reynolds number
$\displaystyle R_e = {\frac{\rho\ a \mid u\mid }{\mu_0}}$\ ,

capillary number $\displaystyle C_a = {\frac{ \mu_0\, \mid u\mid
}{\sigma_2}} $\ ,

 Weber number $\displaystyle
W_e = { \frac{\Delta \rho\, u^2 a }{\sigma_2}}$\ ,

Bond number $\displaystyle B = { \frac{\Delta \rho \, g a^2
}{\sigma_2}} $\ , \newline

\noindent where $\Delta \rho$ is the density difference between the
two fluids, $g$ is the acceleration due to gravity and $\mu_0$ is
the viscosity in the liquid bulk.
\newline \noindent A vertical capillary system treated
with an asymptotic simplification was undertaken by Pukhnachev,
\cite{pukhnachev}, and Baiocci and Pukhnachev, \cite{baiocci}. Their
study involves  Navier-Stokes fluids paired with an imposed apparent
dynamic contact angle as an additional condition at the points of
contact between the free boundary and the solid wall. The problem of
the dissipative function becoming infinite near the contact line is
avoided by introducing a slip length at the solid wall. It appears
that for two-dimensional or axi-symmetric motions, when the
capillary number tends to zero, the free interface tends uniformly
to the equilibrium position.

\noindent Moreover, to minimize the effect of gravity on the
meniscus shape, the capillary tube radii are generally of order less
than $1$ mm.
 A quantitative assessment of relative effects of gravity
and capillary forces can be made on the basis of the Bond number. In
Foister, \cite{foister}, it is experimentally and theoretically
proved that the capillary effects dominate over gravity for all
systems when $B<1$. Likewise, the relative importance of inertial
and capillary effects can be characterized by the Weber number. In
all experimental systems, when $W_{e}<10^{-3}$, inertial effects did
not significantly affect the three phase boundary motion.\newline In
1980, Lowndes, \cite{lowndes}, performed numerical calculations of
the steady motion of the fluid meniscus in a capillary tube. He
showed that when $R_{e}<10^{-2}$, the meniscus formed by an
incompressible Newtonian fluid can be determined by using the
creeping flow approximation. In this case, our proposed
non-Newtonian model has the same streamlines as the Navier-Stokes
model, \cite{gouin}. The method used by Lowndes considers a slip
length less than $10$ Angstr\"{o}ms near the contact line. This
length is of  the same order as the distance from the contact line
where the fluid is no longer Newtonian. Comparisons between the
calculated and observed contact angles are in accordance with the
experiments of Huh and Mason, \cite{huh}; conclusions agree with
those in  part 1. In the partial wetting case, when $C_{a}<10^{-2},$
a Young's angle in the interval $\displaystyle[{\frac{\pi
}{6}},{\frac{5\pi }{6}}]$ verifies Eq. (3).\newline

 \noindent In what follows, we wish to simplify the formidable mathematical problem and
  describe the  dynamics of the air-liquid
interface by a suitable approximation with as few parameters as
possible. The fluids are considered to be Newtonian except in a
close (molecular) vicinity of the contact line, \cite{gouin}.
Boundary condition on the meniscus is the classical
equation (2). For slow movements, when $%
R_{e}<10^{-2},C_{a}<10^{-2},W_{e}<10^{-3}$ and $B<1$, the capillary
flow is axi-symmetric, and the shape of the surface $\Sigma _{t2}$
at a given time $ t $ is described as a spherical cap where the
angle with $\Sigma_{t1}$ is the apparent dynamic contact
angle.\newline \noindent At $20^{\circ }$ Celsius, for water, in
c.g.s. units,  $ \rho =1, \mu _{0}=0.01$ and for an air-water
interface $\sigma _{2}=72$. When $a=0.1$, the above conditions on
$R_e, C_a, W_e$ and $B$ are largely verified with a velocity $u$ of
some centimeters per hour. Similar results are obtained for glycerol
($\rho =1.26,\mu _{0}=11.8$ and $\sigma _{2}=63$) with a velocity
$u$ less than a meter per hour.

\noindent In the developments of  section 4 below, these conditions
are assumed to be fulfilled.

\section{ The  motion of the contact line}

Let us present some analytical formulae related to the geometry of
the contact line, see fig. 1.

\noindent The   volume of the spherical cap with height $h$ and span
$2a$ is
\[
V={\frac{{\pi h}}{6}}\,(h^{2}+3a^{2}) ,\ \ \ \mathrm{where}\ \
h\,\in \,[-a,a].
\]
The volume of the liquid in the cylindrical tube is constant, so

\[
V_{0}=\pi a^{2}y_{\varepsilon }+{\frac{{\pi h}}{6}}\,(h^{2}+3a^{2})
\]
is constant.
 Let  $l_{0}$ be such that $\pi
a^{2}l_{0}=V_{0}$. Then,
$$
y_{\varepsilon }-l_{0}=-{\frac{h}{{\ 6a^{2}}}}\ (h^{2}+3a^{2})\hskip10pt%
\mathrm{and \hskip10pt} \tan\,({\frac{\pi }{4}}-{\frac{\theta
}{2}})=-{\frac{h}{a}}\  . \eqno(4)
$$
If $\displaystyle\lambda = h/a\, $ where $\,\lambda \in \lbrack
-1,1] $, then
$$
\mathrm{cos}\ \theta =-{\frac{{2\lambda }}{{(1+\lambda ^{2})}}}\
  . \eqno(5)
$$
Eqs. (4), (5) establish the connections between $y_{\varepsilon }$
and $h$ and between $y_{\varepsilon }$ and $\theta $. From
$y_{\varepsilon }=z-L$ and taking Eq. (4) into account, we deduce
$$
z=l_{0}+L-{\frac{h}{{\ 6a^{2}}}}\ (h^{2}+3a^{2})\,\equiv
\,l_{0}+L-{\frac{a}{6}}\lambda \,(\lambda ^{2}+3) .\eqno(6)
$$
For a given value of $t$ and for $\lambda $ belonging to $[-1,1]$, $
z$ is a decreasing function of $\lambda $ with values in the
interval $ \displaystyle [\,l_{0}+L-{2a}/{3}, l_{0}+L+ {2a}/{3}
\,]$.

\noindent With $  u = {dz}/{dt}$ and the above expressions, the
dynamic Young-Dupr\'e Eq. (3) yields
\[
\nu\, {\frac{dz}{dt}} = {\frac{2 \, \lambda }{(1 + \lambda^2)}} \,
\sigma_2 - \sigma_1\, ,
\]
where $\nu$ is also a function of $z$ through the Young  angle
$\theta$. Taking Eq. (1) into account, we obtain
$$
\nu\, {\frac{dz}{dt}} = f(z,t)\, +\, K \, \mathrm{sin}\, {\frac{z}{%
\varepsilon\, a}}\ \ {\rm with}\ \   f(z,t) = \,  {\frac{2 \,
\lambda }{(1 + \lambda^2)}} \, \sigma_2 - \sigma_{10} \ \  {\rm and}
\ \ K = k\, \sigma_{10}\ . \eqno (7)
$$
 \textit{Remarks}: The surface of the spherical cap
 is $  \pi \,(h^{2}+a^{2})$. The capillary energy of the
total system per unit  length of the circumference of the capillary
is $W_{e}$; so,
\[
\,2\pi aW_{e} = \,\pi \sigma _{2}\,(h^{2}+a^{2}) +2\pi
a\int_L^{z}\sigma _{1}\ dz ,
\]
Relation (1) implies,
\[
W_{e}=  \sigma _{2}\,(\frac{h^{2}}{2a}+\frac {a}{2}) +
\int_L^{z}\sigma _{10}\ dz - \int_{z_{I}}^{z} k \sigma_{10}\
\mathrm{sin} \,{\frac{z}{\varepsilon a}} \ dz ,
\]
where $z_I$ denotes the abscissa of the point I (see fig. 1).
Consequently,
\[
W_{e}=W_{0}+W_{1}+C^{te}
\]
with
$$
W_{0}=\sigma _{2}\,{\frac{h^{2}}{{2a}}}+\sigma _{10}\,(z-L),\ \
W_{1}=k\,\sigma _{10}\,\varepsilon \,a\,\ \mathrm{cos}\,{\frac{
z}{\varepsilon a}},\eqno(8)
$$
and $C^{te} = \sigma _{2}\, {a}/{2} - k\,\sigma _{10}\,\varepsilon
\,a\,\ \mathrm{cos}\, ({ z_I}/{\varepsilon a})$ is an additional
constant.

\noindent  By taking relation (6) into account, we obtain
 \[
{\frac{\partial h}{\partial z}}=-{\frac{2a^{2}}{{h^{2}+a^{2}}}}\ \
{\rm and} \ \
{\ {\frac{\partial {W_{0}}}{\partial z}}=-f(z,t),\hskip10pt{%
\frac{\partial {W_{1}}}{\partial z}}=-k\,\sigma _{10}\,\mathrm{sin}\,{\frac{z%
}{\varepsilon a}}} \ .
\]
Eq. (7) can now be written in the form
$$
\nu \,{\frac{dz}{dt}}=-{\frac{\partial {W_{e}}}{\partial z}}\eqno(9)
$$
which is  an equation for a unidimensional motion of a mechanism
with linear friction where the effects of inertia (associated with
the second derivative ${{d^{2}z}/{dt^{2}}}$) are absent. In
subsection 4.1, we will see that the behaviour of the solutions of
Eq. (9), when the parameter $\varepsilon$ is vanishingly small, is
completely different from what happens when the effects of inertia
are present in a dynamics equation.

\subsection{ Asymptotic analysis of the motion of the contact line}

The differential equation (9) yields the motion of the contact line.
The behaviour of the solutions when $\varepsilon$ tends to zero has
already been studied in the literature within the framework of a
problem associated with  shape memory alloys (Abeyaratne, Chu and
James, \cite{abeyaratne2}).
\medskip

\noindent Note that Eq. (7) implies  ${\partial f}/{\partial z}\
(z,t) = ({4 \sigma_2 }/{a})\, (\lambda^2-1)/ (\lambda^2 + 1)^3 $;
for $t$ given and $\lambda \in [-1, 1]\, $, $f(z,t)$ is a decreasing
function of $z$. We consider the case associated with an average
surface tension $\sigma_{10}$ of partial wetting corresponding to an
average Young  angle $\theta_m$ such that,
\[
\sigma_2 \,\mathrm{cos}\, \theta_m + \sigma_{10} = 0.
\]
Consequently, the inequality,  $ \mid{ {\sigma_{10}}/{\sigma_2}}
\mid\, < 1 $, is verified. Since $\sigma_2$ is positive, this
relation is equivalent to the inequalities
$$
\sigma_2 + \sigma_{10} > 0 \hskip 10pt \mathrm{and } \hskip 10pt
\sigma_2 - \sigma_{10} > 0 . \eqno (10)
$$
The equation of the contact line motion is only valid for partial
wetting, nevertheless, we obtain the following limit relations
associated with the limit values of $\lambda$:

\noindent Eq. (5) expresses the connection between $\lambda$ and
$\theta$; analytically, for $\lambda = -1$, ${\rm} cos\ \theta = 1$
and the Young  angle $\theta = 0$. Eq. (6) expresses the connection
between $\lambda$ and $z$;  for $\lambda = -1$, $z = l_0 + L + { {2
a }/{3}}$, for which Eq. (7) implies, $f(z,t) = -\sigma_{10} -
\sigma_2 $.

\noindent In the same way, for $\lambda = 1$, the Young angle
$\theta = \pi$ and $z = \displaystyle l_0 + L - { {2 a }/{3}}$, for
which $f(z,t) = - \sigma_{10} + \sigma_2$. \medskip

\noindent Consequently,  under conditions (10), for $K\equiv
k\,\sigma _{10}$ sufficiently small, the differential equation (7)
fulfills exactly the conditions of the \textit{asymptotic theorem}
presented in the Appendix. Thus, it is possible to deduce the
behaviour of the solutions from Eq. (7) when $\varepsilon $ tends to
0. Our aim is not to discuss the general solution of  Eq. (7) when
$L(t)$ is an arbitrary function of $t$. We only consider two
significant cases encountered in experiments (Raphael and de Gennes
\cite{raphael}), namely $(i)$ piston at rest and $(ii)$ piston in
uniform motion ($L(t) = v_0\ t$).
\medskip

\noindent With ${\ y_{\varepsilon }\equiv z-L(t)\equiv l_{0}-(
{a}/{6})\lambda (\lambda ^{2}+3)}$, let us define the function $F$
by
$$
F(y_{\varepsilon })\equiv \,f(z,t)\,\equiv
\,\displaystyle{\frac{2\,\lambda
}{(1+\lambda ^{2})}}\,\sigma _{2}-\sigma _{10}\,\equiv \,-\sigma _{2}\,%
\mathrm{cos}\,\theta -\sigma _{10}.\eqno(11)
$$
Then, $y_{\varepsilon }$ satisfies the relation
\[
\nu\, \frac{dy_{\varepsilon }}{dt}\, {+}\, \nu\
v_{0}=F(y_{\varepsilon })+\,K\,\mathrm{sin}\,{\frac{z}{\varepsilon
\,a}}\ ,
\]
where $y_{\varepsilon }(0)=y_{0}$ and $\nu $ is a function of $%
y_{\varepsilon }$.
 Now, straightforward adaptation of the
\textit{asymptotic theorem} in this simple case yields the
macroscopic behaviour of the contact line:
\newline
When $\varepsilon \rightarrow 0$, $y_{\varepsilon }(t)$ converges
uniformly to\ $y(t) \in C^{1}[\Re ^{+}]$ and satisfies the
differential equation
$$
\nu (y)\,{\frac{dy}{dt}}=G(y)\, ,\eqno(12)
$$
where $y(0)=y_{0}$, $\nu $ is now a function of $y\;$in place of
$y_{\varepsilon }$ and
\[
G(y)=\left\{
\begin{array}{ccl}
\,\,[F^{2}(y)-K^{2}]^{\frac{1}{2}}-\nu v_{0}, & \mathrm{if} &
\,\,\,\,y<y^{-}, \cr\cr    -\, \nu v_{0},\ \ \ \ \ \ \ \ \ \ \ \ \ \
\ \ \ \ \ \ \ & \mathrm{if} & \,\,\,\,y^{-}<y<y^{+},\hskip 2.8cm
(13) \cr\cr\ -\,[F^{2}(y)-K^{2}]^{\frac{1}{2}}-\nu v_{0}, &
\mathrm{if} & \,\,\,\,y^{+}<y\, .
\end{array}
\right.
\]
$y^{-}$ and $y^{+}$ are   constants verifying the relations
\[
F(y^{-})=K\,\,\,\,\mathrm{and}\,\,\,\,F(y^{+})=-K\, .
\]

\noindent Now, we study the two main classes of the dynamical
systems (Penn and Miller, \cite{penn}): those in which the interface
is in non-equilibrium and moves to an equilibrium, and the others in
which the advancing interface is driven on the solid wall of the
cylinder with a constant velocity.

\subsection{${\bf(i)}$  Piston at rest}

In this subsection, the piston is fixed ($v_{0}=0$ and we take
$z\equiv y_{\varepsilon }$).
 In Eq. (8), the potential $W_{0}$ is now independent of $t$  and   a
convex function such as that shown in fig. 2.
\begin{figure}[h]
\begin{center}
\includegraphics[height= 7 cm, width= 10
 cm]{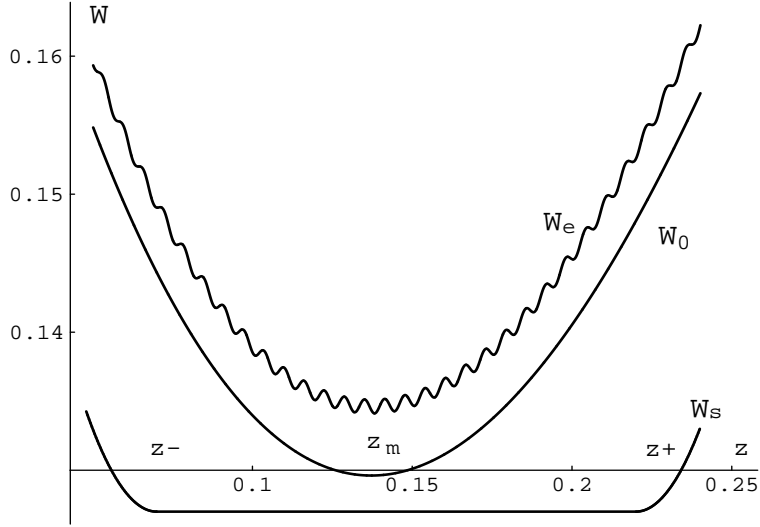}
\end{center}
\caption{ Traces  of the three potentials  $W_e, W_0, W_s$ when the
piston is at rest. In suitable units, we can take the values $
a=1,\,K=0.005,\,\protect\sigma _{10}=-0.5,\,\protect\sigma _{2}=1$.
The curves are centered around $z = z_{m}$ associated with
$\displaystyle \protect\theta_{m}=\mathrm{Arccos}\,0.5$
(coordinate $z$ is defined up to an additive constant). The potentials $%
W_{0}$, $W_{e}$ and $W_{s}$ have been shifted up or down for better
visibility (since potentials are defined up to an additive
constant). Note that $W_{0}$ is the limit of $W_{e}$ when
$\protect\varepsilon $ tends to $0$, but this is not so for $W_{s}$
which is fundamentally different: tangent lines at the limiting
points $z^{-}$ and $z^{+}$ are parallel to the $z$-axis and the function $%
G(z)=-\displaystyle \partial W_{s}/ \partial z$ is of square root
type at these points. } \label{fig2}
\end{figure}
 The differential equation corresponding to the
asymptotic behaviour of Eq. (9) when $\varepsilon \rightarrow 0$
 is expressible in terms of the potential $W_{s}$   as follows
\[
{\frac{\partial W_{s}}{\partial y}}=-G(y)\, ,
\]
where $G(y)$ is defined in $(13)$. When $K$ is sufficiently small,
\textit{the driving force} $G(y)$ has square root singularities (see
(13) with $v_{0}=0$). The functions $W_{0}$ and $W_{s}$ are convex,
but $W_{s}$ is not the limit of $ W_{e}$ when
$\varepsilon\rightarrow 0$ (as $W_{0}$ is). When changing the scales
\textit{- i.e. when} $\, \varepsilon \, $ \textit{tends to zero -}
$\, \, W_s$ cannot be regarded as the limit of the sum of separate
energies associated with the mean energy $W_0$ and the energy of the
perturbation $W_e-W_0$.
 This shows that by changing the scale we lose the
additivity property of the energy for the solution of the limit differential equation: $%
W_{s}$ is not the potential energy limit. However, a physical
interpretation of the previous limit behaviour can be given. When
$\varepsilon \rightarrow 0$, the potential $W_{e}$ admits on $
[y_{\varepsilon }^{-},y_{\varepsilon }^{+}]$ a large number of local
minima whose respective distances converge to zero with $\varepsilon
$. For any initial position $y_{0} \in [y_{\varepsilon
}^{-},y_{\varepsilon }^{+}]$ the closest local minimum is reached in
an infinite time. When $\varepsilon $ tends to zero, any initial
position $y_{0}$ of the interval $[y_{\varepsilon
}^{-},y_{\varepsilon }^{+}]$ is located between two local minima
whose gap tends to zero with $\varepsilon $. On a macroscopic scale
the local minima find themselves together with the initial value
which becomes a \textit{position of equilibrium}. Note that the
angle $\theta _{A}(0)$, given by the value $ \lambda $, drawn from
Eqs. (5), (6), corresponds to the $y_{\varepsilon }^{-}$ -position.
In the same way, an angle $\theta _{R}(0)$ corresponds to the
$y_{\varepsilon }^{+}$-position. \noindent In fact $K\equiv
k\,\sigma _{10}$ is positive and, consequently, $\theta _{A}(0)$ and
$\theta _{R}(0)$ are solutions of the relations
$$
\mathrm{cos}\,\theta _{R}(0)\,=\,{\frac{{-\,\sigma _{10}+k\,\sigma _{10}}}{%
\sigma _{2}}}\,\,\,\,\mathrm{and}\,\,\,\,\mathrm{cos}\,\theta _{A}(0)\,=\,{%
\frac{{-\,\sigma _{10}-k\,\sigma _{10}}}{\sigma _{2}}}\ . \eqno(14)
$$
Then,
\[
\theta _{R}(0)<\theta _{m}<\theta _{A}(0)\ .
\]
An angle $\theta \in [\theta _{R}(0),\theta _{A}(0)]$ corresponds to
an initial position $y_{0}\in \lbrack y_{\varepsilon
}^{-},y_{\varepsilon }^{+}]$. This is not so for $y_{0}\notin
\lbrack y_{\varepsilon }^{-},y_{\varepsilon }^{+}]$. Indeed, in this
case,  the second member $G(y)$ of the asymptotic equation of motion
(12) of the contact line, constitutes a non zero \textit{attractive
force} toward the points $y_{\varepsilon }^{-}$ and $ y_{\varepsilon
}^{+}$. The points $y_{\varepsilon }^{-}$ and $y_{\varepsilon }^{+}$
are now reached in finite time due to the explicit convergence of
the solutions of Eq. (12) (see Appendix for the computation of these
times). \noindent The position $y_{\varepsilon }^{-}$ (resp.
$y_{\varepsilon
}^{+}$) is the \textit{position of equilibrium} associated with $%
y_{0}<y_{\varepsilon }^{-}$ (resp. $y_{0}>y_{\varepsilon }^{+}$) to
which angle $\theta _{A}(0)$ (resp. $\theta _{R}(0)$) corresponds.

\noindent This \textit{static case} indicates that the Young angle
$\theta$ is included in the interval $[\theta_R(0),\theta_A(0)]$.
The final value of $ \theta$, denoted by $\theta_f$, depends on the
initial position $y_0$ of $y$ and, consequently, on the initial
value $\theta_0$ of $\theta$. We obtain the asymptotic behaviour
$$
\left\{
\begin{array}{ccl}
\,\,\theta_0 \in \, (\,\theta_R(0),\theta_A(0)\,) &
{\Longrightarrow} & \, \, \, \, \theta_f = \theta_0\ , \cr \cr
\theta_0 \leq \theta_R(0) & {\Longrightarrow}
& \, \, \, \, \theta_f = \theta_R(0)\ , \cr \cr \theta_0 \geq \theta_A(0) & {%
\Longrightarrow} & \, \, \, \, \theta_f = \theta_A(0)\ .
\end{array}
\right. \eqno (15)
$$

\subsection{{\bf (ii)} Piston in uniform motion}

\noindent When $v_{0}\not=0$, it is easy to prove that for all
initial conditions $y_{0}$, a constant solution to Eq. (12) is
reached in a finite time. In the Appendix, an order of magnitude of
this time is calculated. We obtain $u=v_{0}$. \medskip

When \textit{the piston advances}, $v_{0}=u>0$ and when \textit{the
piston retreats}, $v_{0}=u<0$. The constant solutions of Eq. (12)
are given by
\[
F^{2}(y)-k^{2}\,\sigma _{10}^{2}-\nu ^{2}\,u^{2}=0 \ .
\]
Taking into account the definition of $F$ in relation (11), the
value of the advancing angle $\theta _{A}(u)$ is
$$
\theta _{A}(u)=\ {\rm Arccos}\,\Biggl(\,{\frac{{\,-\,\sigma _{10}-\sqrt{%
k^{2}\,\sigma _{10}^{2}+\nu ^{2}\,u^{2}}\,}}{\sigma _{2}}}\,\Biggr)\
. \eqno(16)_1
$$
The same arguments yield the value of the retreating angle $\theta
_{R}(u)$
$$
\theta _{R}(u)=\ {\rm Arccos}\,\Biggl(\,{\frac{{\ -\ \sigma _{10}+\sqrt{%
k^{2}\,\sigma _{10}^{2}+\nu ^{2}\,u^{2}}\,}}{\sigma _{2}}}\,\Biggr)\
, \eqno(16)_2
$$
and we deduce the inequalities
\[
\theta _{R}(u)<\theta _{R}(0)<\theta _{m}<\theta _{A}(0)<\theta
_{A}(u)\ .
\]
The Young angle $\theta $ is a function of the contact-line celerity
$u$ in an universal form. In fact, the line friction $\nu $ depends
on $\theta $ but $ \nu/\nu _{0}$, (where $\nu _{0}$ is a constant
friction value) belongs to the interval $[1,1.7]$ when  $\theta \in \left[ 30%
{{}^\circ}%
,150%
{{}^\circ}%
\right] $ (see part 1). A simple approximation of (16) consists to
consider $\nu $ as a constant (for example an average value of the
line friction like $1.35\,\,\nu _{0}$).  With such an approximation,
graphs for $\theta _{A}(u)$ and $\theta _{B}(u)$ are associated with
the three values $ \sigma _{10}/\sigma _{2} \equiv
-\mathrm{cos}\,\theta _{m}\,,  k/ \sigma _{2},  \nu /\sigma _{2}$.
The different forms of graphs are presented in fig. 3. They are
drawn from $\theta = 0^\circ$ to $\theta = 180^\circ$ although our
model is no longer valid  for these limit values.

\begin{figure}[h]
\begin{center}
\includegraphics[height= 7 cm, width= 12 cm]{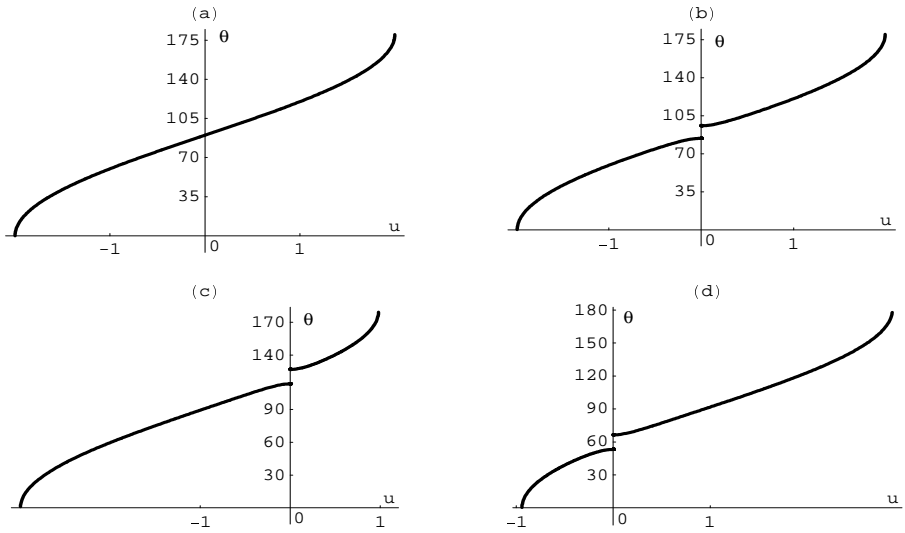}
\end{center}
\caption{Values of the Young angle as functions of the celerity $u$
of
the contact line. We plot different formal cases associated with relations $%
(16)_1$ and $(16)_2$ and convenient values of $\protect\sigma_{10}, \protect%
\sigma_2, k, \protect\nu$. In panel (a), $k = 0$ and the Young angle $%
\protect\theta_m$ is $\displaystyle {\ 90^\circ}$. In panels (b),
(c) and (d), $k \not = 0$ and the Young angle $\displaystyle
\protect\theta_m$ has values $\displaystyle {90^\circ}$,
$\displaystyle {120^\circ}$ or $ \displaystyle    {60^\circ}$. In
cases (b), (c) and (d), we note a hysteresis phenemenon due to the
discontinuity of $\protect\theta$ for $u = 0 $. The unit of the
u-axis depends on the fluids. } \label{fig3}
\end{figure}

\section{ Comparison between results, experimental data and other models
obtained in the literature}

We propose a model of hysteresis of the contact system at the
interface liquid-fluid-solid with the aid of which experimental
results can be interpreted. Moreover, we find that the behaviour
matches that obtained by kinetic molecular arguments in the
literature.

\subsection{The dynamic line tension behaviour}

The differential equation (7) is equivalent to Eq. (3) in the form
$$ \displaystyle \nu\,
{\frac{dz}{dt }} = - \sigma_2\, \mathrm{cos } \, \theta\, -
\sigma_1\ .
$$
Following the expression for the solid-surface energy, the emerging
equation admits a macroscopic behaviour expressed by the
differential equation (12). When the piston is at rest, the
macroscopic contact-line motion is governed by the equation (see Eq.
(12)):
$$
\nu\, {\frac{dz}{dt }} = \pm \, [ (\sigma_2\, \mathrm{cos } \,
\theta\, + \sigma_{10})^2 - k^2 \, \sigma_{10}^2 ] ^ {\frac{1}{2 }}
\ , \eqno (17)
$$
where the sign $+$ or $-$ depends on the direction of the line
motion and $k \, \sigma_{10} \equiv {\max\mid \sigma_1 - \sigma_{10}
\mid} $ is the maximum of the fluctuations in the fluid-solid energy
with respect to its average value $ \overline \sigma_1$.

\noindent The \textit{average value} of the surface tension is $\tau
= \sigma_2\, \mathrm{cos } \, \theta\, + \sigma_{10}$. It has the
dimension of a force per unit length. If $\tau_m = - k \sigma_{10}$
represents the value of $\tau$ at $\theta = \theta_A(0)$ (see
relation (14)), when the angle $\theta$ is close to $\theta_A(0)$,
we obtain
$$
\nu {\frac{dz}{dt }} = (\tau^2 - \tau^2_m)^{\frac{1}{2 }} \approx
\sqrt {2\, \tau_m}\, (\tau - \tau_m)^{\frac{1}{2 }}\ . \eqno (18)
$$
When the Young angles are small, on expanding $\theta$ to order 2,
we obtain
$$
\theta^2 - \theta_A^2(0) \, \equiv \, \beta_0 (\tau - \tau_m)\ ,
\eqno (19)
$$
where $\beta_0$ is a suitable constant.

\noindent The results are extendable to the case of a piston in
advancing motion. Eq. (17) remains unchanged, but the dynamic angle
of contact is such that
\[
(\sigma_2\, \mathrm{cos } \, \theta\, + \sigma_{10})^2 - k^2
\sigma_{10}^2 - \nu^2\, u^2 = 0\ .
\]
Noting that  $\displaystyle  \tau_u = - \sqrt {(\sigma_2\,
\mathrm{cos } \, \theta\, + \sigma_{10})^2 - \nu^2\, u^2 }$, the
previous results are unchanged but $\mid \tau_u \mid < \mid \tau_m
\mid$.

\noindent When the contact line retreats, it is easy to present
similar calculations and to obtain analogous results.
\medskip

\noindent These results are similar to those obtained in the
literature by molecular statistics or single defects (Ruckenstein
and Dunn, \cite{ruckenstein}). For example, Eq. (18) and Eq. (19)
and an analysis of their consequences are also presented in Raphael
and de Gennes, \cite{raphael}, Joanny and Robbins, \cite{joanny}.

\subsection{ Limit velocities of the contact line}

Limit velocities for advancing and retreating contact lines have
been experimentally found and described in detail (Hoffmann,
\cite{hoffman}, Dussan, \cite{dussan2}, Blake and Ruschak,
\cite{blake1}, Chen, Ram\'e and Garoff, \cite{chen}, Decker {\it et
al}, \cite{decker}). These velocities are generally outside the
domain of validity of Eq. (3). Nevertheless, we formally extend our
calculations to the case $\theta \in ( 0, \pi )$, and the asymptotic
behaviour obtained in subsection 4.1 makes it possible to calculate
the limit velocities! Of course, this extension corresponds to the
fact that the forms of the graphs of $\theta$ as a function of $u$
presented in fig. 3 are similar to experimental graphs proposed in
the literature.
 Indeed, the advancing Young angle must be smaller than $\pi$%
. The celerity of the contact line is $u _ \pi$ and the associated
line friction $\nu_\pi$. Relation $(16)_1$ yields
$$
\nu_\pi \, u_ \pi = \sqrt{(\sigma_2 - \sigma_{10})^2 - k^2\,
\sigma_{10}^2 } \ . \eqno (20)
$$
In the same way the retreating Young angle must be larger than $0$.
The celerity of the contact line is now $u_0$ and the associated
line friction $ \nu_0$. Relation $(16)_2$ yields
$$
\nu_0 \, u_0 = - \sqrt{(\sigma_2 + \sigma_{10})^2 - k^2\,
\sigma_{10}^2 } \ . \eqno (21)
$$
Notice that the velocities $u_0$ and $u _ \pi$ do not have the
same absolute value.\\
With this crude approximation, when the line friction is chosen with
a constant value $\nu$,  knowledge of $\sigma_2$, $\sigma_{10} =
\overline \sigma_1$, $u_0$ and $u_ \pi$ allows us
to determine $K = k \, \sigma_{10}$ and $\nu$.\\

In the Appendix, it is proved that the representation (1) for the
surface energy is only a convenient way to consider calculations
with surface heterogeneities. The representation is extendable to
any periodic function with variations on short intervals with
respect to macroscopic sizes. Consequently, the previous results
make it possible to investigate the surface quality and line
friction by simple measurements.

\subsection{ Connection between the dynamic contact angle, the line celerity
and the line friction}

The purpose of this paragraph is  to show by comparison with simple
experiments that our model leads to qualitative and perhaps some
quantitative coincidental behaviours of the contact system with
experimental evidence. The simplest way is to consider the surface
inhomogeneity given by our model represented by relation (1). The
values of $\sigma_{10}$, $\sigma_2$ and $k$, allow us to draw the
graphs of the applications given by $ (16)_1$ and $(16)_2$, (see
fig. 3). Only the ratio $\sigma_{10}/ \sigma_2$ and the value of $k$
are important. \medskip

\noindent We present in fig. 4 experimental layouts drawn in the
literature for an advancing motion of the contact line (Zhou and
Sheng, \cite{Zhou}). The similarity of the theoretical graph and the
experimental data is striking. Nevertheless, we must note that the
experimental data in the literature is always presented in
logarithmic scales.
 This comparison   makes it possible to obtain numerical values for
 the line friction $\nu$ (as
an average) and the limit velocities $u_0$ and $u_\pi$. We use the
experimental results for liquid flows in contact with air in
capillary tubes. Results do not take any explicit account of the
inhomogeneity of the tube walls.
 In the model suggested by relation (1), the solid
surface inhomogeneity is represented by the factor $k$. In fact, $k
\ll 1$, and in relations (20) and (21) the term $k^2 \,
\sigma_{10}^2$ is neglected in comparison to the terms $(\sigma_2 -
\sigma_{10})^2$ and $(\sigma_2 + \sigma_{10})^2$.
\begin{figure}[h]
\begin{center}
\includegraphics[height= 7 cm, width= 12 cm]{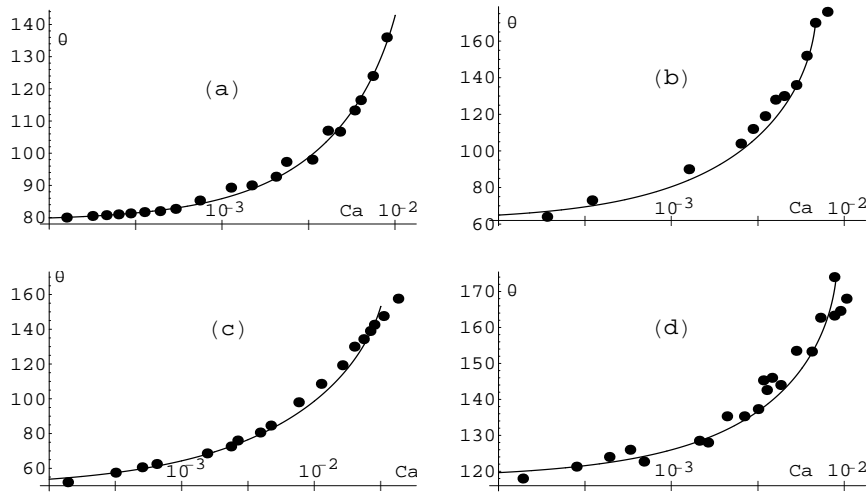}
\end{center}
\caption{Apparent dynamic contact angle  plotted as a function of
$C_a = \protect\mu_0 \, {u / \protect\sigma_2}$. The $C_a$-axis is
based on a logarithmic scale. Panels (a) and (b) are associated with
the data of J.P. Stokes et al. (ref. 13 from Zhou and Sheng,
\cite{Zhou}). Panels (c) and (d) are associated with the data of
G.M. Femigier and P. Jenffer (ref. 12 from Zhou and Sheng,
\cite{Zhou}). Solid curves are calculated with $(16)_1$. The solid
surfaces are assumed to be sufficiently smooth for the hysteresis to
be
small. It is easy to adjust the values of $\protect\sigma_{10}, \protect%
\sigma_2$ and $\protect\nu$ to fit the experimental data given by
points. } \label{fig4}
\end{figure}

\noindent In order to be in agreement with the experimental
conditions, we suppose the diameter of the tube to be about $1 $ mm.
We deduce the numerical relations between the celerities $ C_a$ and
$u$ for two liquids. These expressions are given in c.g.s. units.
For water we obtain $u = 7200\, C_a \, $ and for glycerol $u = 4.2
\, C_a$. The experimental curves yield limit velocities of the
contact line. In various measurements they represent values for
$C_a$ and range between $10^ {-2}$ and $10^ {-1}$. For an
air-glycerol interface, these values correspond to the limit
celerity $u _ \pi$ with a value between $0.042 \, $ cm s $^ {-1}$
and $0.42 \,$ cm s $^ {-1}$, and for an air-water interface the
limit celerity $u _ \pi$ lies between $72 \,$ cm s $^ {-1}$ and $720
\, $ cm s $^ {-1}$. They are widely out of the range of validity of
our model, but for a wonder, they are in good agreement with
experimental results! \newline For glycerol and a limit celerity of
$0.42 \,$ cm s $^ {-1}$ and for a static wetting angle $\theta_A(0)$
of about $50^\circ$, we obtain a line friction of about 240 Poise.
For water and  a limit celerity of $100 \, $ cm s $^ {-1}$, the line
friction is about one Poise. Let us note that the case of water is
obviously less realistic than the case of glycerol. These values of
the line friction are of the same order of magnitude as those
obtained with relation (26) in part 1.

\section{ Conclusion}

We proposed in this paper a dynamic model of slow movements of the
contact system between two fluids and a solid surface. Comparison
between our results and recent experiments or behaviour inferred
from statistical physics shows good agreement on the qualitative
level and, more unexpectedly, on the quantitative side.
 The most significant innovation in this paper is the
introduction of the notion of line friction. This term is essential
to the construction of the model. The line friction depends on the
Young angle, but an order of magnitude for its average value  can be
obtained by using experimental measurements in the literature.

\medskip

\noindent The Young-Dupr\'e relation (3) takes the inhomogeneity on
the solid surface into  account. It describes the microscopic
behaviour of the Young angle. The inhomogeneity is distributed at
distances between a few tens to some hundred Angstr\"oms. This
distance is that of the operating ranges of intermolecular forces
which command the surface energies. At the lower part of this scale,
energies are homogenized and these distances on a macroscopic scale
are no longer significant. The shape of the surface is prescribed on
larger scales.

\medskip

\noindent The results are independent of the radius of the tube.
Indeed, the hysteresis behaviour solely depends on the
physico-chemical properties of the solid surface. The universal form
of the hysteresis loop makes it possible to discuss the general case
independently of any particular apparatus. Relations (20) and (21)
can be written without difficulty using the inhomogeneity of the
solid surface in a form different from Eq. (1).

\noindent Not all the efforts on the contact line  have the same
effects: the one associated with the rapid oscillations of the
surface energy of the wall,  ($ K {\rm sin}\, (z/\varepsilon a)$),
produces a work that tends to zero when the wave length of the
oscillations tends to zero. This effort appears in the macroscopic
expression of the contact line motion represented by Eq. (12). It is
noteworthy that the weak differences between the potentials $W_0$
and $W_e$ have huge implications. This may seem surprising at first
sight. It is due to the fact that the contact line is massless and
consequently its motion equation is not in the same form as for
material systems. Some authors model contact lines as lines with
matter (Slattery, \cite{slattery}); nevertheless, the inertial force
associated with the mass of the line is generally of an order
smaller than the magnitude of the force due to the line friction.

\medskip

\noindent Finally we note that jumps on the inhomogeneity are
considered by Jansons, \cite{jansons}. A shift factor is proposed by
Hoffman, \cite{hoffman}, correcting the relation between $u$ and
$\theta$. These considerations appear unnecessary in our model where
movements are slow and $\theta$ is the apparent dynamic contact
angle. Furthermore, the experimental literature notes the influence
of evaporation on the relaxation time for approaching the apparent
dynamic contact angle (Penn and Miller, \cite{penn}). Similar
inferences are drawn for equipments which are subjected to
vibrations (Marmur, \cite{marmur}). The relaxation time of the
apparent dynamic contact angle is obtained by phenomenological
methods (Hoffman,\cite{hoffman}, Penn and Miller, \cite{penn}). It
corresponds to the calculations carried out in the Appendix.

\noindent Lastly, other models for high-speed motions of the contact
line   using a notion of dynamic surface energy are also considered
in the literature: Shikhmurzaev, \cite{shikhmurzaev}, Blake, Bracke
and Shikhmurzaev, \cite{blake2}. Their study is outside the scope of
our slow movement model.

\section{Appendix: Proof of the fundamental theorem}

Consider the differential equation
$$
\nu (z)\,{\frac{dz}{dt}}=f(z,t)\,+\,K\,sin\,{\frac{z}{\varepsilon }}\,\eqno%
(A1)
$$
\newline
\newline
(we present the same differential equation as Eq. (7), but $f$ may
be a
more general function  than listed in (7), and $a=1$ corresponds to the choice of a convenient unit of length).%
\newline
\newline
The following hypotheses are assumed:

$\bullet\  K$ and $\varepsilon$ ($\varepsilon \ll 1$) are two
strictly positive constants;

$\bullet\ f$ is  continuously differentiable   for any $z \in \Re$,
and for any $t \in \Re^+$,

$\hskip 0.3 cm {\partial f}/{\partial z} < 0$ ;

$\bullet\ $There exist $z^{-}$ and $z^{+}$ belonging to $C^{1}(\Re
)$ such that for any $t \in \Re ^{+}$
$$
 f(z^{-},t)=K\,\,\,\,\mathrm{and}\,\,\,\,f(z^{+},t)=-K\, ;\eqno(A2)
$$

$\bullet\ \nu $ is a strictly positive continuous function of\ $z$.
\medskip

\textit{Remarks:}  The fact that for $t$ fixed, $f(z,t)$ is a
decreasing function on $[0,+\infty) $, implies\ $z^{-}<z^{+}$.
Differential equation (A1) yields a single solution $z_{\varepsilon
}(t)$ defined in $\Re ^{+}$ with\ $z_{\varepsilon }(0)=z_{0}$, (the
differentiable equation fulfills the conditions of uniqueness of the
Cauchy problem  (Hartman, \cite{hartman})).

\noindent We obtain the fundamental result that gives the behaviour of $%
z_{\varepsilon }(t)$ when $\varepsilon $ tends to 0 from the
following theorem
\medskip

\textbf{Asymptotic theorem} \medskip

\noindent When $\varepsilon $ tends to $0$, $z_{\varepsilon }(t)$
converges uniformly to $z(t)$ belonging to $C^{1}[\Re ^{+}]$ and
satisfying the differential equation
$$
\nu (z)\,{\frac{dz}{dt}}=g(z,t)\, ,\eqno(A3)
$$
where\ $z(0)=z_{0}$ and
\[
g(z,t)=\left\{
\begin{array}{ccl}
\,\,[f^{2}-K^{2}]^{\frac{1}{2}} & \mathrm{if} & \;z<z^{-}(t)\,
,\cr\cr0\ \ \ \ \ \ \ \ \ \ &
\mathrm{if} & \,\,z^{-}(t)<z<z^{+}(t)\, ,\cr\cr\displaystyle-[f^{2}-K^{2}]^{%
\frac{1}{2}} & \mathrm{if} & \,\,\,z^{+}(t)<z\,  .
\end{array}
\right.
\]
\medskip The theorem, proved for $\nu (z) =$ constant, is extended without
difficulty for $\nu (z)$ belonging to $[\alpha ,\beta ]$ where
$0<\alpha
<\beta <\infty \,\,$\footnote{%
If $  \xi =\int^z \nu (\varsigma)d\varsigma$, $\xi $ is a strictly
increasing function of $z$ corresponding to a bounded change of
length scale depending on the considered point and $f_{1}(\xi
,t)=f(z,t)$ is a decreasing function of $\xi $.}.

\bigskip

Let us give an elementary proof of the fundamental theorem. For a
complete demonstration using Young measures, we refer to Abeyaratne,
Chu and James, \cite{abeyaratne1}, \cite{abeyaratne2}. We just
consider the case $\nu =$ constant ($\nu =1$ with a convenient unit)
and $f$ independent of $t$. For the variable $z$ belonging to a
compact interval, the segment $[-1,1]$ is considered. The hypothesis
$ {\partial f}/{\partial z} < 0$ corresponds to  strict convexity of
$W_{0}\, \Big (  \partial W_{0}/{\partial z}=-f(z) \Big ) $. For a
given $\varepsilon $, consider $z_{0}\in [-1,1]$ as the initial
value of the solution $z_{\varepsilon }(t)$ of Eq. (A1). The
critical points of  Eq. (A1) are the roots of
\[
f(z)+K\,\mathrm{sin}\,{\frac{z}{\varepsilon }}=0\, .
\]
The roots belong to the interval $[z^{-},z^{+}]$. Let us consider
the case for which $z_{0}$ is smaller than $z^{-}$; when $\,T\,$ is
sufficiently small, it is the same for $z_{\varepsilon }(t)$ with
$t\in \lbrack 0,T]$. Let us calculate the limit values $t$ according
to position $z$ when $\varepsilon $ tends to zero. For $\varepsilon
\ll 1$, the function $\, \mathrm{ sin}\,({{z}/{\varepsilon }})$
oscillates rapidly between -1 and 1. Let  $ z_{0,\varepsilon }$ be
the value immediately above or equal to $z_{0}$ such that $\,
\mathrm{sin}\,({ {z_{0,\varepsilon }}/{\varepsilon }}) =-1$ and let
$z_{1,\varepsilon }$ be the value immediately below or equal to $
z_{\varepsilon }(t)$ such that $\, \mathrm{sin}\,({
{z_{1,\varepsilon }}/{\varepsilon }})=1$. Then,
\[
\lim_{\varepsilon \rightarrow \,0\,\,}\int_{z_{0,\varepsilon
}}^{z_{1,\varepsilon }}{\frac{dz}{{f(z)+K\mathrm{sin}\displaystyle{\frac{z}{%
\varepsilon }}}}}=t\, .
\]
Let us divide the interval $[z_{0,\varepsilon },z_{1,\varepsilon }]$
in intervals of length $\varepsilon \pi $. Then,
\[
\int_{z_{0,\varepsilon }}^{z_{1,\varepsilon }}{\frac{dz}{{f(z)+K\mathrm{sin}%
\displaystyle{\frac{z}{\varepsilon }}}}}=\sum_{p=0}^{N-1}\int_{z_{0,%
\varepsilon }+p\varepsilon \pi }^{z_{0,\varepsilon }+(p+1)\varepsilon \pi }{%
\frac{dz}{{f(z)+K\mathrm{sin}\displaystyle{\frac{z}{\varepsilon
}}}}}\ ,
\]
where  $ N = (z_{1,\varepsilon}-z_{0,\varepsilon})/(\varepsilon \pi)
$

\noindent At each interval $\,[{z_{0,\varepsilon }+p\varepsilon \pi },{%
z_{0,\varepsilon }+(p+1)\varepsilon \pi }]\,$ the change of variables $\,s=%
\mathrm{sin}\,\displaystyle{\frac{z}{\varepsilon }}\,$ yields
\[
\int_{z_{0,\varepsilon }}^{z_{1,\varepsilon }}{\frac{dz}{{f(z)+K\mathrm{sin}%
\displaystyle{\frac{z}{\varepsilon }}}}}=\sum_{p=0}^{N-1}\int_{-1}^{+1}{%
\frac{\varepsilon \,ds}{{(f(\lambda )+Ks)\sqrt{1-s^{2}}}}}\ ,
\]
where $\lambda =\varepsilon \,\mathrm{arcsin}\,s$ and $\lambda \in
\displaystyle{[\varepsilon p\pi -\varepsilon {\frac{\pi
}{2}},\varepsilon p\pi +\varepsilon {\frac{\pi }{2}}]}$.
 Because $f$ is continuous, this expression has the same limit when
$\varepsilon $ tends to zero as
\[
{\frac{1}{\pi }}\sum_{p=0}^{N-1}\varepsilon \pi \int_{-1}^{+1}{\frac{ds}{{%
(f(\varepsilon p\pi -\varepsilon {\frac{\pi
}{2}})+Ks)\sqrt{1-s^{2}}}}}
\]
or
\[
{\frac{1}{\pi }}\sum_{p=0}^{N-1}\varepsilon \pi \int_{-1}^{+1}{\frac{ds}{{%
(f(\varepsilon p\pi +\varepsilon {\frac{\pi
}{2}})+Ks)\sqrt{1-s^{2}}}}}\ ,
\]
which are two sums of Darboux  integrals
\[
\displaystyle{\ \int_{z_{0,\varepsilon }}^{z_{1,\varepsilon
}}{\frac{1}{\pi } }\left(
\int_{-1}^{+1}{\frac{ds}{{(f(z)+Ks)\sqrt{1-s^{2}}}}}\right) \,dz} \,
.
\]
When $\varepsilon $ tends to zero, this expression converges to
\[
\int_{z_{0}}^{z(t)}{\frac{1}{\pi }}\left(
\int_{-1}^{+1}{\frac{ds}{{(f(z)+Ks) \sqrt{1-s^{2}}}}}\right) \,dz=t
\, .
\]
Because $f(z)^{2}\geq K^{2}$, the change of variables $\,s=\mathrm{sin}%
\,\alpha \,$ yields
\[
\int_{-1}^{+1}{\frac{ds}{{(f(z)+Ks)\sqrt{1-s^{2}}}}}={\frac{\pi }{\sqrt{%
f(z)^{2}-K^{2}}}}
\]
and finally,
\[
\int_{z_{0}}^{z(t)}{\frac{dz}{\sqrt{f(z)^{2}-K^{2}}}}=t \ ,
\]
which implies
\[
\dot{z}=\sqrt{f(z)^{2}-K^{2}} \ .
\]
In the same way, for $z_{0}$ above $z^{+}$ we obtain,
\[
\dot{z}=-\sqrt{f(z)^{2}-K^{2}} \ .
\]
When $z_{0}$ belongs to the interval $[z^{-},z^{+}]$, it is easily
seen that there exists a critical point$\;\overline{z_{\varepsilon
}}$ of Eq. (A1) such that $\mid \overline{z_{\varepsilon
}}-z_{0}\mid <b\,\varepsilon $ where $\,b\,$ is a positive constant
depending only on $K$
and $f$. The solution of Eq. (A1) tends in a monotonous way towards $%
\overline{z_{\varepsilon }}$. Consequently, for $t$ belonging to
$[0,+\infty
\lbrack $, $z_{\varepsilon }(t)$ converges uniformly to $z_{0}$ when $%
\varepsilon $ tends to zero. The macroscopic law associated to
$z_{0}$ belonging to $[z^{-},z^{+}]$ is $\dot{z}=0$.

\medskip

\textbf{General oscillations} \medskip

\noindent We consider the superimposed effect of an arbitrary smooth
periodic function $\displaystyle P({ {z}/{\varepsilon}})$ in place
of $ \mathrm{sin} \, ({ {z}/{\varepsilon}})$. We can assume without
loss of generality that $\displaystyle P({ {z}/{\varepsilon}})$ has
a zero average, (in other cases, we add a constant to $f$).

\noindent The amplitude of $P$ gives the placement of the flat
region as in figure 2:
\[
-\mathrm{max}\,P\,\leq \,f(z,t)\,\leq \,-\mathrm{min}\,P
\]
With the same hypotheses as in the theorem we define $z^{-}(t)$ and
$ z^{+}(t) $ such that
\[
f(z^{-}(t),t)=-\mathrm{min}\,P\,\,\,\,\mathrm{and}\,\,\,\,f(z^{+}(t),t)=-%
\mathrm{max}\,P\ ;
\]
so the flat region remains $[z^{-}(t),z^{+}(t)]$.

\noindent In the special case when $f$ is independent of $t$, the
calculation is developed in the same way as previously. The results
are unaltered with $- \mathrm{min} \, P$ in place of $K$ and $-
\mathrm{max} \, P $ in place of $-K$.

\noindent The complete proof of this extension is given in
Abeyaratne, Chu and James, \cite{abeyaratne2}.
\medskip

\textbf{Relaxation time associated with Eq. (A3)} \medskip

\noindent We consider the case when $g$ is explicitly independent of
$t$ and $\nu $ is constant. For $z_{0}<z^{-}$, let
\[
\chi =\nu
\,\int_{z_{0}}^{z^{-}}\,{\frac{du}{\sqrt{f(u)^{2}-K^{2}}}}\ .
\]
For $z_{0}$ near $z^{-}$, $f(u)+K\sim 2\,K$ and $f(u)-K\sim
(u-z^{-})f^{\prime }(z^{-})$. Then, $f(u)^{2}-K^{2}\sim
2\,K(u-z^{-})f^{\prime }(z^{-})$ and
\[
\chi \,\sim \,\nu \,\sqrt{\frac{2}{{-K\,f^{\prime
}(z^{-})}}}\,\sqrt{ z^{-}-z_{0}}\ .
\]
The $\chi $-value yields the magnitude of the relaxation time
necessary to obtain the final position of the contact line.

\section*{Acknowledgments}
 The author would like to express his gratitude to Professor Hutter and the anonymous
 referees for helpful suggestions during the review process.

\end{document}